\documentclass{article}

\usepackage{PRIMEarxiv}
\usepackage[utf8]{inputenc}
\usepackage[T1]{fontenc}
\usepackage{hyperref}
\usepackage{url}
\usepackage{booktabs}
\usepackage{amsfonts}
\usepackage{nicefrac}
\usepackage{microtype}
\usepackage{lipsum}
\usepackage{fancyhdr}
\usepackage{graphicx}
\usepackage{listings}
\graphicspath{{media/}}
\usepackage{caption}
\usepackage{subcaption}
\usepackage{amsmath}

\title{Long Short-Term Memory Pattern Recognition in Currency Trading}

\author{
  Jai Pal \\
  Independent Researcher \\
  \texttt{jaipal9621@gmail.com}
}

\begin{document}
\maketitle

\begin{abstract}
This study delves into the analysis of financial markets through the lens of Wyckoff Phases, a framework devised by Richard D. Wyckoff in the early 20th century. Focusing on the accumulation pattern within the Wyckoff framework, the research explores the phases of trading range and secondary test, elucidating their significance in understanding market dynamics and identifying potential trading opportunities. By dissecting the intricacies of these phases, the study sheds light on the creation of liquidity through market structure, offering insights into how traders can leverage this knowledge to anticipate price movements and make informed decisions. The effective detection and analysis of Wyckoff patterns necessitate robust computational models capable of processing complex market data, with spatial data best analyzed using Convolutional Neural Networks (CNNs) and temporal data through Long Short-Term Memory (LSTM) models. The creation of training data involves the generation of swing points, representing significant market movements, and filler points, introducing noise and enhancing model generalization. Activation functions, such as the sigmoid function, play a crucial role in determining the output behavior of neural network models. The results of the study demonstrate the remarkable efficacy of deep learning models in detecting Wyckoff patterns within financial data, underscoring their potential for enhancing pattern recognition and analysis in financial markets. In conclusion, the study highlights the transformative potential of AI-driven approaches in financial analysis and trading strategies, with the integration of AI technologies shaping the future of trading and investment practices.
\end{abstract}

\section{Introduction}
The study of financial markets involves navigating a complex landscape where myriad factors influence asset prices and market dynamics. Traders and analysts often rely on various tools and techniques to interpret market behavior and identify trading opportunities. Among these methodologies, the Wyckoff Phases stand out as a framework that provides insights into the cyclical nature of market movements and the interplay between supply and demand.

Developed by Richard D. Wyckoff in the early 20th century, the Wyckoff Phases offer a systematic approach to analyzing market trends and identifying potential turning points. The framework consists of several distinct phases, each reflecting a specific stage in the evolution of market sentiment and investor behavior. In this research, we focus on one of the key patterns within the Wyckoff framework: the accumulation pattern.

\section{Wyckoff Pattern}
In the context of Wyckoff analysis, the phases of trading range and secondary test play pivotal roles in understanding market dynamics and identifying potential trading opportunities.

During the trading range phase, the market enters a period of consolidation and indecision, where prices fluctuate within a relatively narrow range. This phase reflects a balance between buyers and sellers as they engage in a tug-of-war for control. The formation of lower lows and lower highs within the trading range signifies a potential shift in market sentiment, with downward pressure gradually diminishing.

Following the trading range, the secondary test phase emerges as a crucial stage in the accumulation pattern. This phase involves a retest of previous support levels, typically accompanied by a decrease in selling pressure and an uptick in buying interest. The creation of liquidity through uniform and equal lows reinforces the bullish bias, signaling a strengthening of underlying demand. This sets the stage for a potential breakout to the upside, as market participants seek to capitalize on the emerging bullish momentum.

Liquidity creation plays a vital role in both phases, facilitating smoother price movements and providing opportunities for market participants to enter or exit positions with minimal impact on prices. By understanding how liquidity is created through market structure, traders can better anticipate price movements and identify optimal entry and exit points in the market. This underscores the importance of incorporating Wyckoff principles into trading strategies, as they offer valuable insights into market psychology and price action dynamics.

\begin{figure}[ht]
	\centering
	\includegraphics[width=0.7\linewidth]{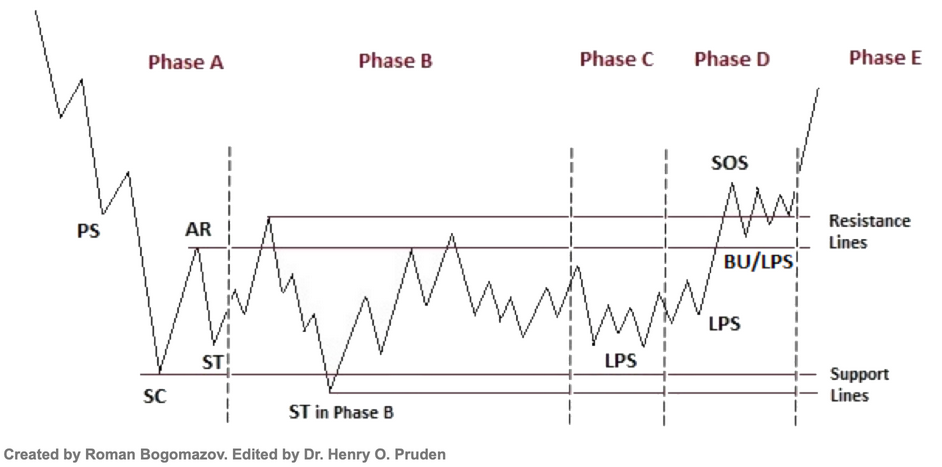}
	\caption[]{Wyckoff Accumulation Pattern}
	\label{fig:accvisual}
\end{figure}

\begin{figure}[ht]
	\centering
	\includegraphics[width=0.7\linewidth]{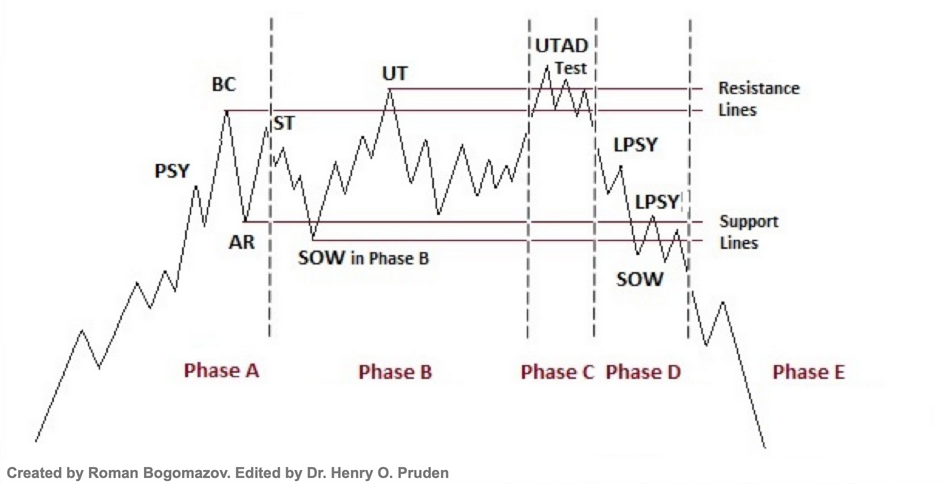}
	\caption{Wyckoff Distribution Pattern}
	\label{fig:distvisual}
\end{figure}

\subsection{Trading Range Phase}
The trading range phase represents a period of consolidation and indecision in the market. During this phase, prices oscillate within a relatively narrow range as buyers and sellers engage in a tug-of-war for control. The formation of lower lows and lower highs within the trading range signals a potential shift in market sentiment, with downward pressure gradually waning.

\subsection{Secondary Test Phase}
Following the trading range, the secondary test phase emerges as a critical juncture in the accumulation pattern. This phase is characterized by a retest of previous support levels, often accompanied by diminishing selling pressure and increased buying interest. The creation of liquidity through uniform and equal lows reinforces the bullish bias, setting the stage for a potential breakout.

\section{Model Selection}
The effective detection and analysis of Wyckoff patterns require robust computational models capable of processing and interpreting complex market data. In this regard, the selection of an appropriate model hinges on the intrinsic characteristics of the data under consideration.
\subsection{Spatial Data and Convolutional Neural Networks}
CNNs operate by using convolutional layers to extract features from input data. These layers consist of filters that slide over the input data, performing element-wise multiplications and summations to produce feature maps. By applying multiple convolutional layers with non-linear activation functions such as ReLU (Rectified Linear Unit), CNNs can learn hierarchical representations of data, capturing increasingly complex patterns and structures.

CNNs excel at processing spatial data due to their inherent ability to exploit local correlations and spatial dependencies within the data. For example, in image recognition tasks, CNNs can identify objects based on the arrangement of pixels and the presence of certain visual features. Similarly, in financial analysis, CNNs can analyze spatial patterns in market data, such as the configuration of candlestick patterns or the distribution of technical indicators, to make predictions or extract meaningful insights.

However, it's important to note that while CNNs are effective for analyzing spatial data, financial market data such as price movements is fundamentally sequential rather than spatial. Unlike images or spatial datasets where the arrangement of elements carries intrinsic meaning, price data represents a temporal sequence of events. Each data point in a price chart is not independent but instead influenced by preceding data points and contextual factors. Therefore, while CNNs can be applied to spatial aspects of financial data, sequential models such as Long Short-Term Memory (LSTM) networks are better suited for analyzing temporal relationships and sequential patterns in price data.

\begin{figure}[ht]
	\centering
	\includegraphics[width=0.7\linewidth]{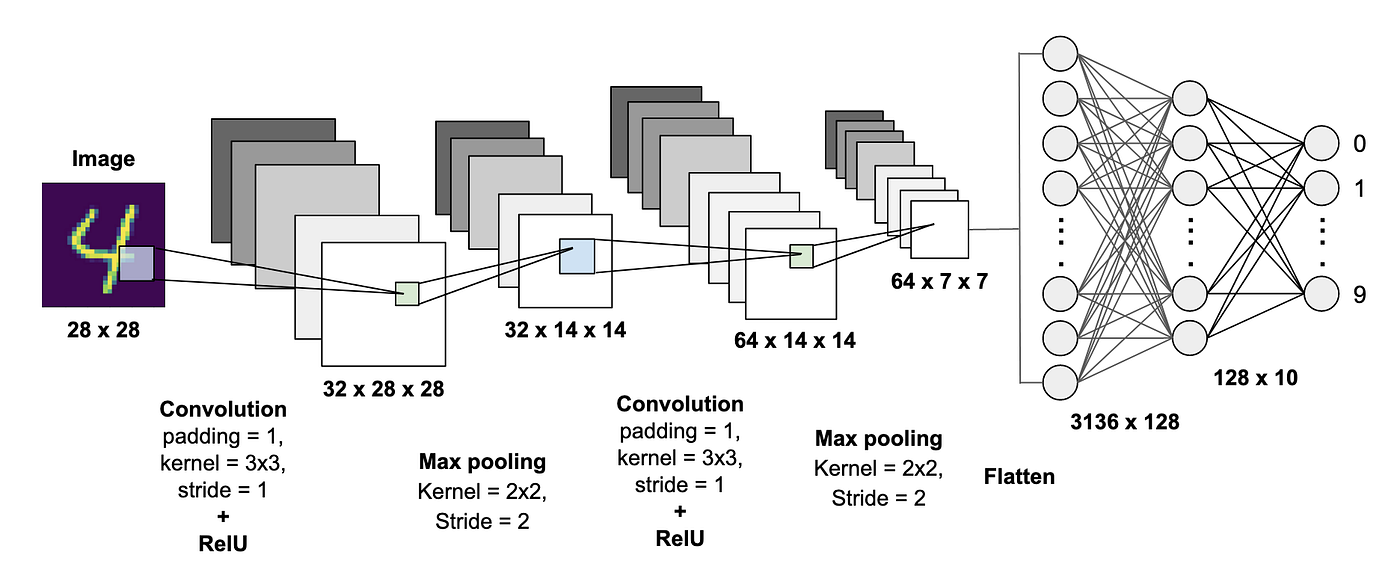}
	\caption{Convolutional Neural Network Diagram}
	\label{fig:cnnmodel}
\end{figure}

\subsection{Temporal Data and Long Short-Term Memory Models}
Temporal data, characterized by sequential information and time-dependent patterns, is ubiquitous in financial markets. Long Short-Term Memory (LSTM) models, a type of recurrent neural network (RNN), excel at modeling sequential data and capturing long-term dependencies. LSTMs are specifically designed to address the challenges of modeling sequential data by incorporating mechanisms for capturing both short-term and long-term dependencies. 

At the core of an LSTM unit are three gates: the input gate, the forget gate, and the output gate. These gates control the flow of information within the LSTM cell, allowing it to selectively retain or discard information over time. The input gate regulates the flow of new information into the LSTM cell, deciding which information to incorporate into the current state based on the input data and the current state of the cell. The forget gate determines which information to discard from the current state, considering both the input data and the previous state of the cell to decide which information is no longer relevant or should be forgotten. Meanwhile, the output gate controls the flow of information from the current state to the output of the LSTM cell, regulating the information that gets passed on to subsequent cells in the network or used for making predictions. 

Unlike traditional RNNs, LSTMs include a cell state that serves as a conveyor belt, allowing information to persist and flow through multiple time steps, thereby enabling them to capture long-term dependencies in the data. By leveraging both short-term and long-term memory mechanisms, LSTMs are capable of learning complex patterns and capturing dependencies that span across the entire sequence of data. In the context of financial markets, where historical price movements and trends play a crucial role in decision-making, LSTMs offer powerful tools for analyzing and forecasting market dynamics.

\begin{figure}[ht]
	\centering
	\includegraphics[width=0.7\linewidth]{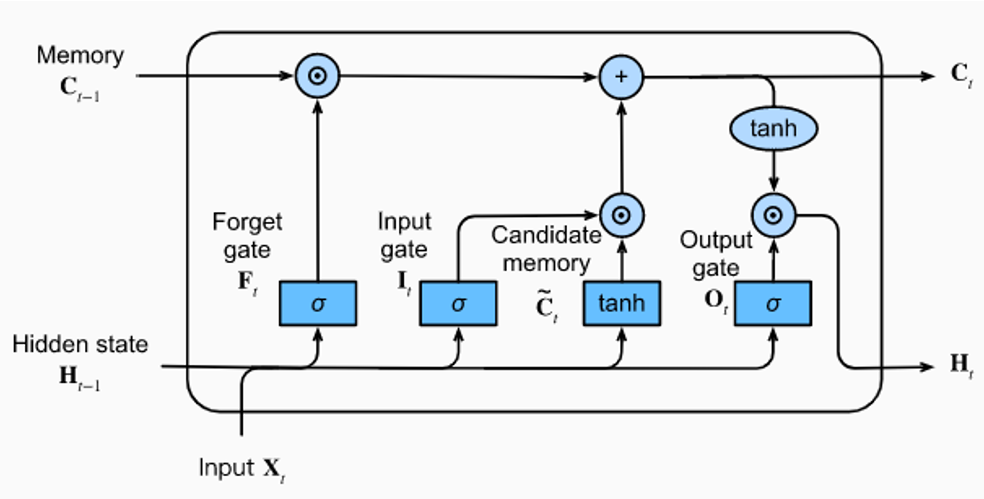}
	\caption{Long Short-Term Memory Model Diagram}
	\label{fig:lstmmodel}
\end{figure}
\clearpage
\section{Data Creation}
The creation of training data for modeling Wyckoff patterns involves two main components: swing points and filler points. Swing points represent significant market movements, while filler points introduce noise and enhance model generalization.
\subsection{Swing Points}
Swing points play a pivotal role in technical analysis, particularly within the framework of Wyckoff methodology. These points represent significant extremas in price movement, capturing moments where the market experiences notable shifts in direction. In the context of Wyckoff analysis, swing points are characterized by the formation of lower lows and lower highs, signaling a potential change in trend direction.

The identification of swing points involves scrutinizing price charts and identifying key price levels where significant highs and lows occur. A lower low is established when a subsequent price low is lower than the preceding low, indicating a downward movement in price. Conversely, a lower high occurs when a subsequent price high is lower than the preceding high, suggesting a downward reversal in price momentum.

By pinpointing these swing points, traders can discern the underlying trends within the market. Lower lows and lower highs signify a downtrend, indicating a prevailing bearish sentiment in the market. Conversely, higher highs and higher lows denote an uptrend, reflecting a bullish bias among market participants.

Understanding the concept of swing points enables traders to identify potential trend reversals and anticipate future price movements. By analyzing the sequence of swing points over time, traders can gain insights into the market's overall direction and make informed trading decisions. Additionally, swing points serve as reference points for defining support and resistance levels, aiding in the identification of key price zones where buying and selling pressure may intensify.

\begin{lstlisting}[language=Python, caption={Creating the Trading Range Phase Pattern}]
	def generate_pattern(validity):
		if validity:
			p1 = random.uniform(0, 100)
			p2 = random.uniform(0, p1)
			p3 = random.uniform(p2, p1)
			p4 = random.uniform(p2, p3)
			return [validity, p1, p2, p3, p4]
		if not validity:
			pattern = [validity, random.uniform(0,100),
			 random.uniform(0,100), random.uniform(0,100),
			  random.uniform(0,100)]
		if pattern[1] > pattern[2] and pattern[2] < pattern[3]
		 and pattern[4] < pattern [3] and pattern[3] < pattern[1]
	 	 and pattern[4] > pattern[2]:
			pattern.remove(pattern[0])
			pattern.insert(0, True)
	return pattern
\end{lstlisting}

\begin{lstlisting}[language=Python, caption={Creating the Secondary Test Phase Pattern}]
	def generate_pattern(validity):
		if validity:
			p1 = random.uniform(0, 100)
			p2 = random.uniform(0, p1)
			p3 = random.gauss(p2, 5) 
			p4 = random.gauss(p2, 5)
			p3 = min(max(0, p3), p1)
			p4 = min(max(0, p4), p1)
			pattern1 = [validity, p1, p2]
			pattern2 = upFiller([p3,p4], p1)
			final_pattern = filler(pattern1 + pattern2)
		return final_pattern
\end{lstlisting}

\subsection{Filler Data}
Filler points are introduced to augment the dataset with additional variability and simulate real-world market conditions. These points are randomly inserted between swing points, creating a more diverse and representative dataset for training the model. By incorporating filler points, the model becomes more robust and capable of generalizing to unseen market scenarios. 

In machine learning, the generation of training data plays a crucial role in the performance and generalization ability of models. Noise generation, specifically the introduction of filler points, is an essential aspect of this process. While swing points represent significant market movements and provide valuable information for training, filler points serve to add variability and randomness to the data, mimicking the noisy nature of real-world datasets.

The importance of noise generation lies in its ability to improve the robustness and generalization of machine learning models. By exposing models to a diverse range of data points, including those that may not directly contribute to the target outcome, filler points help prevent overfitting and ensure that models can effectively capture underlying patterns in the data.

Moreover, noise generation allows models to learn to distinguish between signal and noise, enabling them to focus on relevant features while disregarding irrelevant fluctuations or anomalies in the data. This process enhances the model's ability to make accurate predictions on unseen data by promoting a more nuanced understanding of the underlying relationships within the dataset.

In the context of modeling Wyckoff patterns, the inclusion of filler points alongside swing points ensures that the model can effectively discern between genuine pattern formations and random market fluctuations. By exposing the model to a diverse range of scenarios, including noise generated by filler points, the model becomes more robust and capable of identifying patterns in real-world market data, thereby enhancing its predictive performance and utility in trading applications.

\begin{lstlisting}[language=Python, caption={Creating Filler Data}]
	def filler(pattern):
		new_pattern = [pattern[0]] 
		num_fillers = 2
		for i in range(1, len(pattern) - 1):
		new_pattern.append(pattern[i])  
	
		for _ in range(num_fillers):
			filler_val = random.uniform(pattern[i], pattern[i+1])
			new_pattern.append(filler_val)
		new_pattern.append(pattern[-1])
		return new_pattern
	
	def upFiller(pattern, upperLimit):
		new_pattern = []
	
		for i in range(len(pattern) - 1):
			start_val = pattern[i]
			new_pattern.append(start_val)
		end_val = pattern[i + 1]
		num_fillers = 1
	
		for _ in range(num_fillers):
			filler_val = random.uniform(start_val, upperLimit)
			new_pattern.append(filler_val)
	
		return new_pattern
	
	

\end{lstlisting}

\section{Activation}
While there exist several other activation functions, such as ReLU (Rectified Linear Unit), tanh (Hyperbolic Tangent), and softmax, the sigmoid activation function remains a popular choice for binary classification tasks due to several key advantages.

One notable advantage of the sigmoid function is its smooth and differentiable nature across its entire domain, which enables efficient gradient-based optimization during model training. This property is particularly advantageous in training deep neural networks, where the smoothness of the activation function helps mitigate the issue of vanishing gradients, allowing for more stable and effective learning.

Moreover, the sigmoid function outputs probabilities that are intuitively interpretable, ranging from 0 to 1. This makes it easier for traders and analysts to understand the model's predictions as likelihoods of a certain outcome, such as the presence of a Wyckoff pattern. The probabilistic nature of the sigmoid activation also lends itself well to uncertainty quantification, allowing traders to assess the confidence level of the model's predictions and adjust their trading strategies accordingly.

Additionally, the sigmoid function naturally handles imbalanced datasets commonly encountered in financial markets, where instances of Wyckoff patterns may be relatively rare compared to non-pattern instances. By outputting probabilities, the sigmoid function provides a continuous measure of the model's confidence in its predictions, allowing for more nuanced decision-making, such as adjusting the threshold for classifying patterns based on the prevailing market conditions.
Overall, while alternative activation functions like ReLU and tanh may offer computational advantages or better performance in certain scenarios, the sigmoid activation function remains a robust choice for binary classification tasks in financial markets due to its interpretability, smoothness, and probabilistic output nature.

\begin{figure}[ht]
	\centering
	\includegraphics[width=0.45\linewidth]{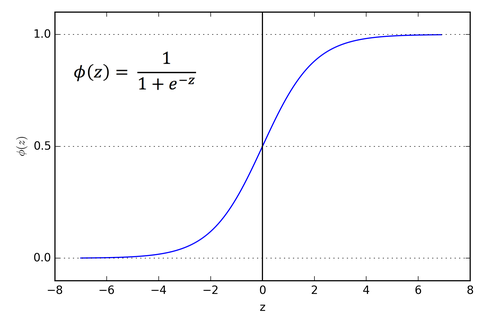}
	\caption{Sigmoid Activation Function}
	\label{fig:sigmoid}
\end{figure}

\begin{lstlisting}[language=Python, caption={Creating the Model with Sigmoid Activation}]
model = Sequential([
LSTM(64, input_shape=(1, n_features)),  # LSTM layer with 64 units
Dense(1, activation='sigmoid')          # Output layer with sigmoid activation 
])
\end{lstlisting}

\section{Results}
In the context of analyzing 100,000 valid and 100,000 invalid Wyckoff patterns, the results of our study showcase the robustness and reliability of deep learning models for pattern recognition tasks in financial markets. Across both phases of the Wyckoff pattern – the trading range phase and the secondary test phase – our models demonstrated exceptional accuracy and minimal loss, highlighting their effectiveness in detecting subtle yet significant patterns within complex financial data.

The model designed to identify the trading range phase yielded impressive results, with a test loss of 0.0207 and a test accuracy of 99.34\%. This high level of accuracy indicates the model's ability to accurately discern the initial phase of Wyckoff patterns, which is characterized by distinctive price movements signaling potential shifts in market trends. By successfully identifying these patterns, the model provides invaluable insights for traders aiming to capitalize on emerging market trends and capitalize on lucrative opportunities.

\begin{figure}[ht]
	\centering
	\includegraphics[width=0.6\linewidth]{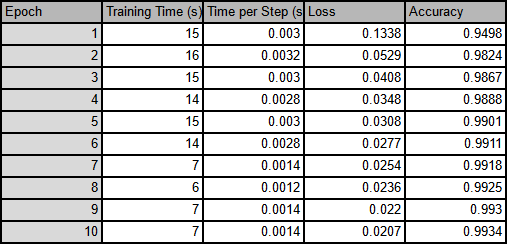}
	\caption{TR Model Accuracy by Epoch}
	\label{fig:tracc}
\end{figure}

Similarly, the model dedicated to detecting the secondary test phase exhibited outstanding performance, achieving a test loss of 0.0010 and a test accuracy of 99.98\%. This exceptional accuracy underscores the model's proficiency in identifying liquidity creation patterns within financial data, a critical aspect of Wyckoff patterns. By accurately capturing these patterns, the model empowers traders to anticipate potential market movements and make well-informed decisions regarding their investment strategies.

\begin{figure}[ht]
	\centering
	\includegraphics[width=0.6\linewidth]{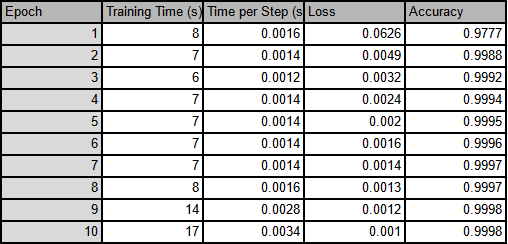}
	\caption{ST Model Accuracy by Epoch}
	\label{fig:stacc}
\end{figure}

Overall, the results of our study underscore the significant potential of AI-driven approaches in enhancing pattern recognition and analysis in financial markets. By leveraging deep learning models, traders and analysts can gain deeper insights into market dynamics, identify emerging trends, and make data-driven decisions that maximize profitability and mitigate risks. As such, our findings contribute to advancing the field of algorithmic trading and highlight the transformative impact of AI technologies on the finance industry.
\begin{figure}[ht]
	\centering
	\includegraphics[width=0.4\linewidth]{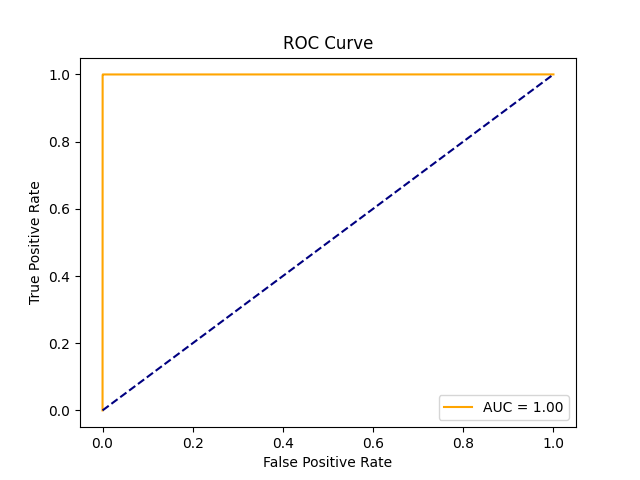}
	\caption{Receiver Operating Characteristics Curve for TR and ST Models}
	\label{fig:stroc}
\end{figure}

\section{Shortcomings}
While the results of this study are promising, it is essential to acknowledge certain limitations and areas for future research. One notable limitation is the reliance on a fixed time step for model training, which may restrict the models' ability to capture nuanced patterns present in larger time frames. Future research could explore adaptive time step approaches or alternative architectures to overcome this limitation and enhance the models' ability to analyze data across various timeframes.

Additionally, the current study focuses primarily on the detection of Wyckoff patterns within financial data. Future research could extend this framework to incorporate other technical indicators, fundamental data, or sentiment analysis to develop more comprehensive trading strategies. Furthermore, exploring the application of reinforcement learning techniques could enable the models to dynamically adapt their strategies based on real-time market conditions, further enhancing their performance and robustness.

\section{Application}
The applications of this research extend beyond academic inquiry, holding significant implications for the fintech industry and algorithmic trading. By leveraging AI-driven models to detect Wyckoff patterns, financial institutions can enhance their trading strategies, optimize investment decisions, and mitigate risks effectively. These models can be integrated into existing trading platforms to provide traders with real-time insights and decision support, empowering them to make data-driven investment choices.

Furthermore, the adoption of AI-driven approaches in algorithmic trading can revolutionize traditional trading strategies, enabling firms to automate trading processes, minimize human error, and capitalize on emerging market opportunities more efficiently. As the financial industry continues to evolve, the integration of AI technologies will play a crucial role in shaping the future of trading and investment practices.

\section{Conclusion}
In conclusion, this study underscores the transformative potential of AI-driven models in pattern recognition and analysis within financial markets. By accurately identifying key phases of Wyckoff patterns with high precision, these models provide valuable insights for traders and investors, enabling them to navigate complex market conditions and capitalize on emerging trends. As we continue to explore the intersection of AI and finance, we unlock new opportunities for innovation, efficiency, and profitability, reshaping the landscape of financial analysis and trading strategies.
\clearpage
\section{Bibliography}
1.17. Neural network models (supervised). (n.d.). Scikit-Learn. Retrieved February 22, 2024, from https://scikit-learn.org/stable/modules/neural\_networks\_supervised.html

API Documentation. (n.d.). TensorFlow. Retrieved February 22, 2024, from https://www.tensorflow.org/api\_docs

Matplotlib documentation — Matplotlib 3.8.3 documentation. (n.d.). Retrieved February 22, 2024, from https://matplotlib.org/stable/index.html

NumPy documentation. (n.d.). Retrieved February 22, 2024, from https://numpy.org/doc/

seaborn: Statistical data visualization — seaborn 0.13.2 documentation. (n.d.). Retrieved February 22, 2024, from https://seaborn.pydata.org/

User Guide — pandas 2.2.0 documentation. (n.d.). Retrieved February 22, 2024, from https://pandas.pydata.org/docs/user\_guide/index.html

(N.d.-a). Retrieved February 22, 2024, from https://miro.medium.com/v2/resize:fit:1400/1*CnNorCR4Zdq7pVchdsRGyw.png

(N.d.-b). Retrieved February 22, 2024, from https://school.stockcharts.com/lib/exe/fetch.php?media=market\_analysis:
the\_wyckoff\_method:wyckoffaccumulation2.png

(N.d.-c). Retrieved February 22, 2024, from https://school.stockcharts.com/lib/exe/fetch.php?media=market\_analysis:
the\_wyckoff\_method:wyckoffdistribution1.png

(N.d.-d). Retrieved February 22, 2024, from https://miro.medium.com/v2/resize:fit:984/1*Mb\_L\_slY9rjMr8-IADHvwg.png

(N.d.-e). Retrieved February 22, 2024, from https://miro.medium.com/v2/resize:fit:970/1*Xu7B5y9gp0iL5ooBj7LtWw.png

\end{document}